\documentclass[aps,twocolumn,showpacs,amsmath,amssymb]{revtex4-1}
\usepackage{graphicx}
\usepackage{dcolumn}
\usepackage{bm}
\usepackage{hyperref}

\begin{document}

\title{Dynamics of local quantum uncertainty among cavity-reservoir qubits}
\author{Mazhar Ali}
\affiliation{Department of Electrical Engineering, Faculty of Engineering, Islamic University Madinah, 107 Madinah, Saudi Arabia \\ 
Email: mazharaliawan@yahoo.com}


\begin{abstract}
We study dynamics of local quantum uncertainty (LQU) for a system of two cavities and two reservoirs. In the start, the cavities treated as 
two qubits are quantum correlated with each other, whereas reservoirs are neither correlated with each other nor with cavities. We answer 
two main questions in this work. First, how local quantum uncertainty decays from two quantum correlated cavities and grows among reservoirs. 
The second question is the examination of LQU developed among four qubits and also shed some light on its dynamics. 
We observe that LQU develops among reservoirs as kind of mirror image to its decay from cavities. For four qubits, we propose how to compute 
LQU such that the method is intuitive and conformable to the observation. We find that among four qubits LQU starts growing from zero to maximum 
value and then decays again to zero as the asymptotic state of cavities is completely transferred to reservoirs. We suggest the experimental 
setup to implement our results. 
\end{abstract}


\maketitle

The study of quantum correlations has attracted considerable interest as it was demonstrated that they can play key role in 
future technologies exploiting this feature of quantum mechanics \cite{Nielsen-Book, Hayashi-Book}. 
One major difference between quantum and classical states is the fact that any local measurements on one part of either bipartite or 
multipartite quantum states leads to uncertainty in outcomes. This randomness is an inherent feature of quantum states. 
Certain quantities like entanglement, non-locality, quantum discord, etc are only defined for quantum states and these might have 
nonzero values for them. 
These quantities have no classical counterpart and in fact classical states can be characterized as those state which have these measures as 
zero \cite{Ollivier-PRL88-2001, Henderson-JPA34-2001, Modi-RMP84-2012}. 
One such measure of quantumness is quantum discord which is defined as the difference between quantum mutual information and classical 
correlations \cite{Ollivier-PRL88-2001, Henderson-JPA34-2001, Modi-RMP84-2012,Luo-PRA77-2008, Ali-PRA81-2010}. This measure is quite hard to 
compute and analytical results are known only for some restricted families of quantum states \cite{Ali-JPA43-2010, Rau-2017}. 
There are some proposals of this measure for multipartite systems as 
well \cite{Mahdian-EPJD-2012, Xu-PLA-2013, Bai-PRA88-2013, Daoud-2012, Buscemi-PRA-2013}. 
Some authors have suggested related measures of quantumness like quantum work deficit \cite{Oppenheim-PRL89-2002}, quantum 
deficit \cite{Rajagopal-PRA66-2002}, measurement-induced non-locality \cite{Luo-PRL106-2011}, etc (see references in \cite{Bera-RPP81-2018}). 
The applications of these measures are found in remote state preparation \cite{Dakic-Nat-2012}, 
entanglement distribution \cite{Streltsov-PRL108-2012, Chuan-PRL109-2012}, transmission of correlations \cite{Streltsov-PRL111-2013}, and 
quantum meteorology \cite{Modi-PRX1-2011}, etc. One fact is common in all these measures that it is not easy to compute 
them for an arbitrary initial quantum state \cite{Horodecki-RMP-2009, gtreview, Chiara-RPP2018}. 
Another measure of quantumness, known as local quantum uncertainty (LQU) has been proposed \cite{Girolami-PRL110-2013}. 
This measure is quantified via skew information which is achievable on a single local measurement \cite{Luo-PRL91-2003}.   
This measure has a closed formula calculated for $2 \otimes d$ bipartite quantum systems. The effects of decoherence on 
this measure is also studied \cite{Wu-AP-2018, Slaoui-2019}. Recently, we have proposed one possible way to compute local quantum uncertainty 
for multi-qubit system \cite{Ali-arx}. 

In this paper, we study quantum correlated two cavity photons interacting with two independent reservoirs. The presence or absence of a photon 
in a cavity defines one qubit, whereas no photon or normalized collective state with a single excitation in the reservoir defines the second qubit. 
Two such quantum correlated cavities interacting with two independent reservoirs effectively forms a system of four qubits. Some interesting aspects 
of quantum entanglement for this system have been investigated before \cite{Lopez-PRL101-2008, Ali-PRA90-2014}. 
Here, we use this system to study dynamics of LQU among different qubits. The most obvious interest is to study its dynamics among 
cavity-cavity qubits, reservoir-reservoir qubits, and among system of all four qubits. This last option brings us to address the question of 
LQU for four-qubit system. As so far LQU is defined and computed for qubit-qudit system only, so first we discuss the extension of this measure 
for multi-qubit system. We stress here that our proposal is not a claim about to compute a kind of "genuine" LQU, nevertheless, it is quite 
intuitive definition and gives reasonable results as we show below. For brevity, we only take couple of examples to demonstrate the main idea and 
results. We find that LQU starts decaying immediately after the interaction from cavity-cavity qubits and reaches a value of zero only at infinity. 
At the same time, it starts growing among reservoir-reservoir qubits in reserve fashion, that is, in a kind of mirror image. The four-qubit LQU 
naturally starts from a zero value, reached to a maximum and then again starts decaying asymptotically.

We briefly describe our physical model and equation of dynamics as follows. Consider the qubits as two uncoupled cavity modes with up-to one photon. Each 
mode interacts independently with its own reservoir. The interaction Hamiltonian (with $\hbar = 1$), for a single cavity mode and a $N$-mode 
reservoir can be written as \cite{Lopez-PRL101-2008}
\begin{eqnarray}
\hat{H} = \omega \, \hat{a}^\dagger \hat{a} + \sum_{k = 1}^N \omega_k \, \hat{b}^\dagger \hat{b} 
+ \sum_{k = 1}^N g_k \big( \hat{a} \hat{b}_k^\dagger + \hat{b}_k \hat{a}^\dagger \big)\,,
\label{Eq:Hamil}
\end{eqnarray}
where $\hat{a}(\hat{a}^\dagger)$ is annihilation (creation) operator for cavity mode, $\hat{b}(\hat{b}^\dagger)$ is same corresponding operator 
for reservoir mode. The first terms describes the energy of cavity photon, second term shows energy of reservoir in any of "N" mode, and third 
term depicts the interaction between cavity modes and reservoir modes. Under the initial conditions of a single photon cavity mode and reservoir 
in the vacuum mode, the combined quantum state can be written as
\begin{eqnarray}
|\psi(0)\rangle_{cr} = |1 \rangle_c \otimes |\bar{{\bf 0}}\rangle_r, \label{Eq:IS}  
\end{eqnarray}
where $ |\bar{{\bf 0}}\rangle_r = \Pi_{k=1}^N |{\bf 0}_k\rangle_r$ is the collective vacuum state of $N$-modes of reservoir $r$. 
The Hamiltonian (\ref{Eq:Hamil}) leads interaction among cavity and reservoir, and in the interaction picture, time evolved state can be written 
as \cite{Lopez-PRL101-2008, Ali-PRA90-2014} 
\begin{eqnarray}
|\psi(t)\rangle_{cr} = \xi(t) |1\rangle_c \, |\bar{{\bf 0}}\rangle_r + \sum_{k=1}^N \lambda_k(t) \, |0\rangle_c \, |{\bf 1}_k\rangle_r \, ,
\label{Eq:TE}
\end{eqnarray}
where the reservoir state $|{\bf 1}_k\rangle_r$ describes the presence of a single photon in mode $k$. The probability amplitude 
$\xi(t)$ converges to $\xi(t) = e^{- \kappa t/2}$ in the limit $N \to \infty$. Eq.(\ref{Eq:TE}) can be written as
\begin{eqnarray}
|\psi(t)\rangle_{cr} = \xi(t) |1\rangle_c \, |\bar{{\bf 0}}\rangle_r + \chi(t) |0\rangle_c \, |\bar{{\bf 1}} \rangle_r \, ,
\label{Eq:TEES}
\end{eqnarray}
where $|\bar{{\bf 1}} \rangle_r = (1/\chi(t)) \sum_{k=1}^N \lambda_k(t) |{\bf 1}_k\rangle$, and the probability amplitude $\chi(t)$ 
converges to $\chi(t) = \sqrt{1-e^{- \kappa t}}$ for large $N$. Eq.(\ref{Eq:TEES}) describes an effective two-qubit 
system \cite{Lopez-PRL101-2008}. We have utilized this equation to obtain a combined time-evolved density matrix for four-qubits for an 
arbitrary state of two-qubit (cavity-cavity) with two reservoirs both in vacuum modes. We have obtained the density matrices of cavity-cavity qubits, 
reservoir-reservoir qubits, or density matrix for any group of qubits, by taking standard procedure of taking partial trace over qubits, which we want 
to exclude from system of interest. The mathematical equations and general solutions for these systems of our interest are quite tedious to present 
here but they are simple and straight forward to obtain. We can generalize such studies for any number of cavity qubits with corresponding reservoirs 
in vacuum modes. With the general solution of the system in hand, we now turn to local quantum uncertainty. 

Local quantum uncertainty (LQU) has recently been defined for $2 \otimes d$ quantum systems \cite{Girolami-PRL110-2013}. This measure detects quantum 
correlations in a given state by applying local measurements. It is similar to quantum discord with the advantage that 
one need not to carry out minimization over von Neumann measurements. To compute LQU, it is sufficient to obtain the maximum 
eigenvalue of a symmetric $3 \times 3$ matrix, which is relatively easy task. 
LQU is defined as the minimum skew information which is obtained via local measurement on qubit part only, that is,  
\begin{equation}
\mathcal{Q}(\rho) \equiv \, \min_{L_A} \, \mathcal{I} (\rho , L_A \otimes \mathbb{I}_B) \,, 
\end{equation}
where $L_A$ is a hermitian operator (local observable) on subsystem $A$, and $\mathcal{I}$ is the skew information \cite{Luo-PRL91-2003} of the density 
operator $\rho$, defined as 
\begin{equation}
\mathcal{I} (\rho , L_A \otimes \mathbb{I}_B) \, = \,- \frac{1}{2} \, \rm{Tr} ( \, [ \, \sqrt{\rho}, \, L_A \otimes \mathbb{I}_B ]^2 \, ) \,.
\end{equation}
It has been shown \cite{Girolami-PRL110-2013} that for $2 \otimes d$ quantum systems, the compact formula for LQU is given as
\begin{equation}
\mathcal{Q}(\rho) = 1 - \rm{max} \, \{ \lambda_1 \,, \lambda_2 \, , \lambda_3 \, \}\,, 
\end{equation}
where $\lambda_i$ are the eigenvalues of $3 \times 3$ symmetric matrix $\mathcal{M}$. The matrix elements of symmetric matrix $\mathcal{M}$ 
are calculated by the relationship
\begin{equation}
 m_{ij} \equiv \rm{Tr} \, \big\{ \, \sqrt{\rho} \, (\sigma_i \otimes \mathbb{I}_B) \, \sqrt{\rho} \, (\sigma_j \otimes \mathbb{I}_B) \, \big\}\,, 
\end{equation}
where $\sigma_i$ are three standard Pauli matrices.

It is tempting to generalize this bipartite definition of LQU to multi-qubit systems. We can consider the multi-qubit system as $2 \otimes D$ system, 
where $D = 2 \otimes 2 \ldots \otimes 2$ represent the remaining $(N-1)$ qubits. It is well known that the multi-qubit systems have 
much richer structure with respect to hierarchy of correlations as compared with bipartite systems. It is possible that some bipartitions 
might be quantum correlated, some other as classically correlated, and some others may be even completely uncorrelated with any other qubits. 
Therefore, it is natural to incorporate every bipartition to account for quantifying LQU for multi-qubit systems. This suggests the application of 
local measurements across each bipartition to detect any quantum correlations and include its outcome in the final result. to this aim, we 
consider $\rho$ as an arbitrary density matrix for $N$ qubits. We apply local measurements on each of 
the qubit $A$, $B$, $\ldots$, $N$, by taking all other qubits as $D = 2^{N-1}$-dimensional systems. In each case, we obtain a symmetric matrix, 
so in total we get $N$ symmetric matrices. For each bipartition, the matrix elements belonging to these $N$ symmetric matrices are calculated 
according to relations 
\begin{eqnarray}
\tilde{m}_{ij}^A =  \rm{Tr} \, \{ \, \sqrt{\rho} \, (\sigma_i \otimes \mathbb{I}_2 \ldots \otimes \mathbb{I}_2)  
\, \sqrt{\rho} \, (\sigma_j \otimes \mathbb{I}_2 \ldots \otimes \mathbb{I}_2) \, \}\,, \nonumber \\ 
\tilde{m}_{ij}^B = \rm{Tr} \, \{ \, \sqrt{\rho} \, (\mathbb{I}_2 \otimes \sigma_i \ldots \otimes \mathbb{I}_2) 
\, \sqrt{\rho} \, (\mathbb{I}_2 \otimes \sigma_j \ldots \otimes \mathbb{I}_2) \, \}\,, \nonumber \\
\vdots \qquad \vdots \qquad \vdots \qquad \vdots \qquad \vdots \qquad \vdots \qquad \vdots \qquad \vdots 
\qquad \vdots \qquad \vdots \qquad \vdots \nonumber \\
\tilde{m}_{ij}^N = \rm{Tr} \, \{ \, \sqrt{\rho} \, (\mathbb{I}_2 \otimes \mathbb{I}_2 \ldots \otimes \sigma_i) \, 
\sqrt{\rho} \, (\mathbb{I}_2 \otimes \mathbb{I}_2 \ldots \otimes \sigma_j) \}\,. 
\end{eqnarray}
The corresponding eigenvalues of such $3 \times 3$ symmetric matrices $\mathcal{\tilde{M}}_i$ can be determined easily. 
The local quantum uncertainties related with each bipartition are defined as follows
\begin{eqnarray}
\mathcal{Q}_{A/BC\ldots N}(\rho) = 1 - \rm{max} \, \{ \text{Spectrum of } \mathcal{\tilde{M}}_A \, \}\,\nonumber \\ 
\mathcal{Q}_{B/AC\ldots N}(\rho) = 1 - \rm{max} \, \{ \text{Spectrum of } \mathcal{\tilde{M}}_B \, \}\,\nonumber \\ 
\vdots \qquad  \qquad \vdots \qquad \qquad \qquad \vdots \qquad \qquad \qquad \vdots \nonumber \\
\mathcal{Q}_{N/ABC\ldots N-1}(\rho) = 1 - \rm{max} \, \{ \text{Spectrum of } \mathcal{\tilde{M}}_N \, \}\,.
\end{eqnarray}
Finally we propose multi-qubit LQU to be defined as
\begin{equation}
\mathcal{Q}(\rho_{N}) = \bigg(\prod_{i = A}^N \, \mathcal{Q}_{i/N_i}\bigg)^{1/N}\,, 
\end{equation}
where $N_i$ are the remaining $N-1$ qubits except $i$. We stress here an important distinction regarding terminology to avoid confusion. 
We propose this multi-qubit LQU with the property that if any number of qubits are either classically correlated or completely uncorrelated 
with others then this measure will be zero. It will be strictly positive if and only if all qubits are somehow quantum correlated with 
each other. However, this does not necessarily mean that our measure quantifies "genuine" quantum correlations like "genuine entanglement" as it is known 
that for multi-qubits there are quantum states which are having negative partial transpose w.r.t. each bipartition (means entangled) but they 
are not genuinely entangled states.     

We now take two specific quantum states for cavity-cavity qubits and study how LQU decays from these cavity qubits and develops among 
reservoir qubits and also among all four qubits. First we consider cavity-cavity qubits to be in the pure state
\begin{eqnarray}
|\psi(0)\rangle =  \, \alpha \, |0 \, 0 \rangle \, + \, \beta \, |1 \, 1\rangle \, . 
\label{Eq:JIS}
\end{eqnarray}
In earlier studies \cite{Lopez-PRL101-2008, Ali-PRA90-2014}, it was shown that quantum entanglement among reservoirs may appear either exactly 
at the same time as it is completely lost from cavities provided $\beta = 2 \, \alpha$, at a later time for $\beta > 2 \, \alpha$ thus giving 
a window where neither cavities nor reservoirs are entangled, or even entanglement may appear among reservoirs earlier than it is lost from cavities for 
$\beta < 2 \, \alpha$. In this regard, we first examine the range of parameter for which it is known that entanglement decays asymptotically. 

\begin{figure}[h]
\scalebox{1.85}{\includegraphics[width=1.85in]{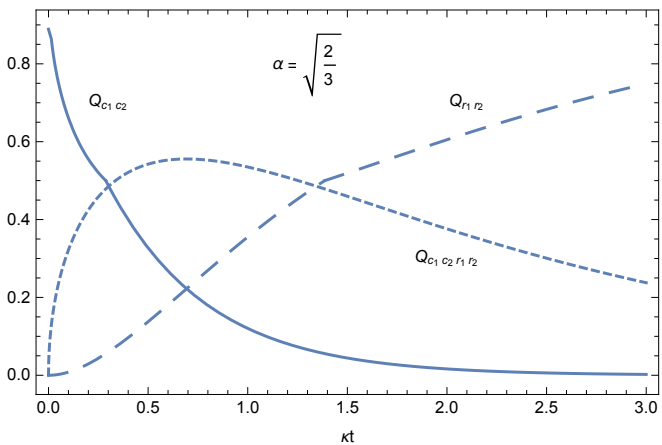}}
\centering
\caption{Quantum local uncertainties are plotted against parameter $\kappa t$ for cavity-cavity qubits (solid line), reservoir-reservoir 
qubits (dashed line) and all four qubits (dotted line) for $\alpha = \sqrt{2/3}$. See text for explanations.}
\label{FIG:CR1}
\end{figure}
Figure (\ref{FIG:CR1}) depicts LQUs plotted against parameter $\kappa t$ with initial condition $\alpha = \sqrt{2/3}$. LQU for cavity-cavity qubits 
$Q_{c_1 c_2}$ is shown by solid line which decays asymptotically similar to entanglement. The dashed curve is for LQU $Q_{r_1 r_2}$ of 
reservoir-reservoir qubits. It is interesting to note that as cavity-cavity LQU decays, corresponding correlations start growing among 
reservoir-reservoir qubits. 
The dotted curve denotes $Q_{c_1 c_2 r_1 r_2}$ which is LQU of combined state of four qubits. We observe that all four qubits get quantum correlated 
immediately as the interaction starts. This curve achieves some maximum value and then starts decaying asymptotically. All this behavior is quite 
intuitive as at the start the correlations are only among cavity-cavity qubits, and at infinity only among reservoir-reservoir qubits. Therefore it is 
natural for any four-partite measure of quantum correlations to start from zero, reach maximum value and then decay to zero. 

\begin{figure}[h]
\scalebox{1.85}{\includegraphics[width=1.85in]{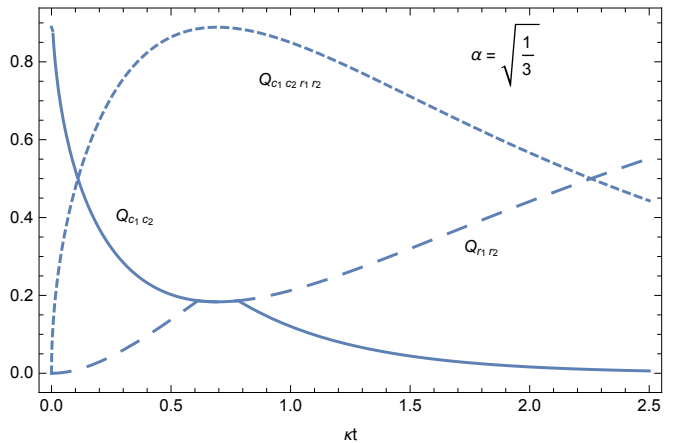}}
\centering
\caption{Same caption as Figure (\ref{FIG:CR1}) except for $\alpha = \sqrt{1/3}$. See text for explanations.}
\label{FIG:CR2}
\end{figure}

Next, we consider the initial condition $\alpha = \sqrt{1/3}$ which leads to sudden death of entanglement in cavity-cavity qubits and sudden birth of 
entanglement among reservoir-reservoir qubits \cite{Lopez-PRL101-2008, Ali-PRA90-2014}. Figure (\ref{FIG:CR2}) shows LQUs for various partitions 
against $\kappa t$. The solid line denotes LQU of cavity-cavity qubits, whereas dashed line is LQU for reservoir-reservoir qubits. 
The dotted line denote LQU for combined state of four-qubits. In this case, we observe an interesting pattern of dynamics of LQU for cavity and 
reservoir qubits. 
As both curves meet together at $\kappa t \approx 0.61$ then curiously they merge together for some time before separating at $ \kappa t \approx 0.82$ 
and follow their subsequent trajectories, cavity LQU decaying and reservoir LQU increasing asymptotically. The four qubit correlation starts from zero 
to maximum value around this time window and then starts decaying asymptotically as well.    

\begin{figure}[h]
\scalebox{1.85}{\includegraphics[width=1.85in]{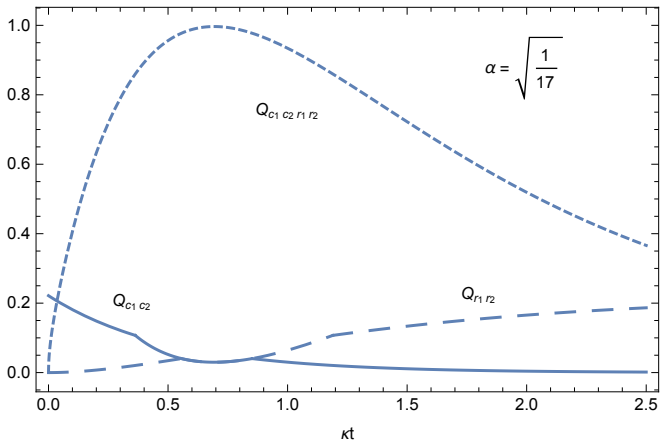}}
\centering
\caption{Same caption as Figure (\ref{FIG:CR1}) except for $\alpha = \sqrt{1/17}$. See text for explanations.}
\label{FIG:CR3}
\end{figure}

We now consider the initial condition which leads to a time window where neither cavity qubits nor reservoir qubits are entangled \cite{Ali-PRA90-2014}. 
One possible choice is by taking $\alpha = \sqrt{1/17}$. Figure (\ref{FIG:CR3}) depicts local quantum uncertainties for cavity-cavity qubits (solid line), 
reservoir-reservoir qubits (dashed line), and four-qubits (dotted line). 
Once again we see the similar behavior as observed in Figure (\ref{FIG:CR2}), however more pronounced. We can see that LQU of cavity qubits (solid line) 
comes to contact with LQU of reservoir qubits at $\kappa t \approx 0.53$, both curves maintain same numerical value up to 
$\kappa t \approx 0.88$. After that curious overlap these curves follow their expected trajectories, one decaying and other increasing asymptotically. 
The four qubit LQU starts from zero, reaches the maximum value around this overlap time and then finally starts decaying asymptotically. 

\begin{figure}[h]
\scalebox{1.85}{\includegraphics[width=1.85in]{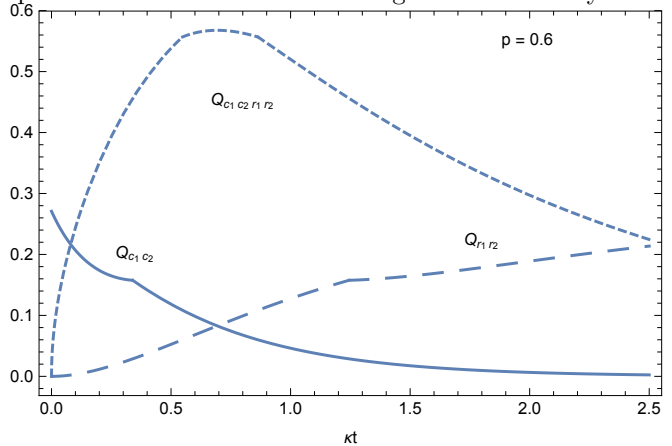}}
\centering
\caption{Local quantum uncertainties for cavity qubits, reservoir qubits, and all four qubits are plotted against parameter $\kappa t$. 
See the text for details.}
\label{FIG:CR4}
\end{figure}

The second specific example we take is the well known Werner states for two qubits, which are single parameter class of states, given as 
\begin{eqnarray}
\rho_p = p \, |\Phi \rangle\langle\Phi| + \frac{(1-p)}{4} \mathbb{I}_4 \, , \label{Eq:8}
\end{eqnarray}
where $p \in [0,1]$ and $|\Phi\rangle = 1/\sqrt{2} (|0,0\rangle + |1,1\rangle)$ is the maximally entangled pure 
Bell state. It is well known that Werner states are entangled for $p \in (1/3,1]$ and separable for $p \leq 1/3$. It is also known that there is always 
sudden death of entanglement except for $p = 1$ and also corresponding sudden birth of entanglement in cavity-reservoir qubits 
setup \cite{Lopez-PRL101-2008, Ali-PRA90-2014}. In order to see, how LQU behaves for this family 
of states, we first take parameter $ p = 0.6$. Figure (\ref{FIG:CR4}) shows local quantum uncertainties for cavity-cavity qubits (solid line), 
reservoir-reservoir qubits (dashed line), and four-qubits (dotted line). As we can see that LQU decays among cavity qubits and start growing among 
reservoir qubits both asymptotically. Four qubit LQU takes its expected trajectory of starting with zero, reaching a maximum value and then decaying 
asymptotically. 

We have studied dynamics of local quantum uncertainty for system of cavity and reservoirs setup. We have found that quantum correlations initially 
present among cavity-cavity qubits are developed among all other partitions. The interesting observation is development of local quantum uncertainty 
among reservoir-reservoir qubits which is kind of mirror image to that of decay from cavity-cavity qubits. We have demonstrated the behavior by taking 
few examples of well known specific quantum states. We have been able to quantify local quantum uncertainty for four-qubit system and also have studied 
its dynamics. Our model and dynamics is very specific with cavities as completely uncorrelated with reservoirs at the start of dynamics and also at 
the end of dynamics as well, because the state at infinity is also a tensor product state between cavities and reservoirs. 
At $\kappa t = \infty$, the joint state factors into $\rho(t = \infty) = | 00\rangle_{c_1 \, c_2}\langle 00| \otimes \rho_{r_1 \, r_2}$ with all 
quantum correlations transferred to the reservoirs. Therefore, intuitively, 
it is clear that any four party quantifier for local quantum uncertainty must have zero value at the start and end of dynamics. This is precisely 
what we have observed in our study. In addition, our generalization of local quantum uncertainty for multi-qubit is straight forward and very simple 
as it does not any any complicated minimization/maximization procedure. The predictions discussed in this work can be easily implemented in experiments 
in optical setups where cavity qubits can be encoded in polarization states and reservoirs with interferometers which couples the polarization to 
path of each photon. Our predictions in Ref.\cite{Ali-PRA90-2014} were already successfully implemented with this scheme \cite{Aguilar-PRL113-2014}. 
Exactly same setup can be used to study local quantum uncertainty as well.   

\end{document}